\documentstyle[12pt]{article}
\begin{document}
 \rightline{SU-ITP-902\ \ \ \ \  }
%\rightline{29 September 1991}

\newcommand{\Psl}{\not\!\! P}
\newcommand{\dsl}{\not\! \partial}
\newcommand{\half}{\frac{1}{2}}
\def\a{\alpha}
\def\b{\beta}
\def\al{\aleph}
\def\g{\gamma}\def\G{\Gamma}
\def\d{\delta}\def\D{\Delta}
\def\e{\epsilon}
\def\et{\eta}
\def\z{\zeta}
\def\t{\theta}\def\T{\Theta}
\def\l{\lambda}\def\L{\Lambda}
\def\m{\mu}
\def\f{\phi}\def\F{\Phi}
\def\n{\nu}
\def\p{\psi}\def\P{\Psi}
\def\r{\rho}
\def\s{\sigma}\def\S{\Sigma}
\def\ta{\tau}
\def\x{\chi}
\def\o{\omega}\def\O{\Omega}
\def\lagr{{\cal L}}
\def\cd{{\cal D}}
\def\k{\kappa}
\def\n{\nabla}
\def\tz{\tilde z}
\def\tF{\tilde F}
\def\ri {\rightarrow}
\def\cf{{\cal F}}

%\begin{titlepage}
\vskip 2.5 cm
\begin{center}
{\large\bf D=10 \  SUPERSYMMETRIC
\vskip 1 cm

  CHERN-SIMONS GAUGE THEORY }\footnote{An invited talk at the
conference ``Strings and Symmetries'', Stony Brook, 20--24 May 1991.}
\vskip 1.2 cm
{\bf R. Kallosh\footnote{On leave of
absence from: Lebedev Physical Institute, Moscow, 117924, USSR }}
\vskip 0.2cm
Physics Department, Stanford University\\
Stanford   CA  94305\\[0.3cm]
\end{center}
\vskip 1.6  cm
\begin{center}
{\bf ABSTRACT}
\end{center}
\begin{quote}

\ \ \ \ \ The  Chern-Simons  ten-dimensional manifestly supersymmetric
non-Abelian gauge theory is presented by performing the second
quantization of the  superparticle theory.

The equation of motion is $F = (d + A)^2 = 0$, where $d$ is the nilpotent
fermionic  BRST operator of the first quantized theory and $A$ is the
anti-commuting  connection  for the gauge group. This equation can be
derived as the condition of the gauge independence of the first quantized
theory in a background field $A$, or from the string field theory
Lagrangian of the Chern-Simons type. The trivial solutions of the
cohomology are the gauge symmetries, the non-trivial solution is given by
the D=10 superspace, describing the super Yang-Mills theory on shell.

\end{quote}

%\end{titlepage}

\section{Introduction}

The ten-dimensional supersymmetric gauge theory has many interesting
properties. It is the largest  globally supersymmetric gauge theory. Its
spectrum coincides with the  spectrum of (non-gravitational) zero modes
of heterotic string theory. The result of its dimensional reduction to four
dimensions (N=4 supersymmetric Yang-Mills theory) is the first example
of a finite four-dimensional  quantum field theory. Understanding the
ten-dimensional  supersymmetric gauge theory would be an important
step toward understanding the geometry of superstring theory and would
give insights into the structure of the  string field theory.

However, despite many attempts,  the ten-dimensional
supersymmetric Yang-Mills theory (SYM)   was not  formulated in a
manifestly supersymmetric, Lorentz invariant and gauge invariant  way.
This is a major challenge to those who are looking for the manifestly
supersymmetric formulation of the  string theory.
\vskip 0.6 cm

 The on shell SYM theory is known in the superspace  $ (X^\mu , \t^\a)$,
where $\t^\a$ is a  ten-dimensional 16-component Majorana-Weyl spinor.
The geometry of this superspace has been constructed by Nilsson
\cite{NIL}.  The gauge connections $ A_{A} = (A_\mu , A_\a)$ depend on
the coordinates of the superspace and take values in some gauge  group
$G$.  The gauge covariant derivatives of the  ten-dimensional superspace
$\n_{A}  = ( \n _{\mu}, \n_{\a} ) $ transform homogeneously under the
Yang-Mills transformation, $\n^{\prime}_{A} =  e^{\L }\n_{A} e^{-\L }$,
 where $\exp {\L(X,\t )}$ is an element of the gauge group $G$.
The anti-commutator of spinorial derivatives $\{ \n_\a, \n _\beta \} =  2
\n_{\a\beta } + F_{\a\beta}$ defines the vector derivative $\n_{\a\beta} =
\g^\mu _{\a\beta} \n_{\mu} $  in case that the spinorial curvature is
constrained to be equal to zero, \begin{equation}
 F_{\a\beta} = 0 \ .
\label{CON}\end{equation}
This is the Nilsson's constraint for ten-dimensional SYM. From this
constraint and from Bianchi identities one can derive the manifestly
supersymmetric gauge covariant Dirac equation  for a ten-dimensional
Majorana-Weyl spinorial superfield $\Psi ^\a (z)$, where
\begin{equation} \g^\mu _{\a \beta} \Psi^{\beta}(X,\t) = [\n_\a ,   \n^\mu ]
\ . \label{EMC2}
\end{equation}
The superfield Dirac equation  contains all equations of motion of the
ten-dimensional supersymmetric Yang-Mills theory. They can be
obtained  from the superfield equation by expanding in $\t $ at  $\t  = 0$.
The superfield Dirac equation is supersymmetric, Lorentz  covariant and
gauge covariant. However, this is not enough for construction of the
corresponding quantum field theory; one still needs to have a manifestly
supersymmetric, Lorentz covariant and gauge covariant Lagrangian.
\vskip 0.6 cm

The purpose of this talk is to present the formulation of the geometry and
the gauge symmetries of the ten-dimensional supersymmetric
non-Abelian gauge theory starting with the first quantized superparticle
theory. The theory which we obtain has a Chern-Simons (CS) structure; it
has manifest supersymmetry, Lorentz symmetry and gauge  symmetry.
The geometric equation of motion reproduces the on shell superspace
describing this theory.

This theory will be constructed by using the first quantized
superparticle theory \cite{BKV} -- \cite{SU}. The second  quantization
will be performed in accordance with the general program developed for
string field theory \cite{SFT}. The way towards a consistent first
quantization of manifestly supersymmetric particle theory was long and
difficult \cite{BM}. Recently two groups claimed to have found a
consistent quantization of the superparticle theory, \cite{BKV} --
\cite{SU}  and \cite{SB}. Different versions of the theory have been used
by these two groups. In what  follows we  are going  to use our version
\cite{K}. This theory of superparticle has the property that
ten-dimensional $N =1$ supersymmetry is realized in a form of a broken
twisted  supersymmetry $N =2(p+1)$,  where $p=0,1,\dots ,\infty $.  In
the light-cone gauge it coincides  with that of the Brink-Schwarz
superparticle.
 \vskip 0.6 cm

The paper is organized as follows:  In Sec.2  we present the general first
quantized path integral in an arbitrary background, by introducing the
gauge covariant BRST operator $\O(P,Q) + A(Q)$. We derive the condition
of the quantum gauge independence  of this functional, which turns out to
be the nilpotency condition of the covariant BRST operator. This
condition, which can be written as an equation of motion of the
Chern-Simons type, is equivalent to the condition of flatness of the target
space.   In Sec.3  we describe the D=10 superspace, which is suggested by
the  classical action of the superparticle \cite{K}. This superspace has
twisted $N=2(p+1)$ supersymmetry. We introduce the analogs of  chiral
(antichiral) subspaces. The chiral  (antichiral) superfields realize the
representations of  $N=1$ supersymmetry, since  chiral  (antichiral)
constraints do not commute with all supersymmetries except one. In Sec.4
we describe the first quantized superparticle theory. We first consider
the background functional with the unconstrained connection and the set
of non-Abelian gauge symmetries. The flatness condition of the target
space geometry will be shown to provide the gauge independence of the
first quantized theory. In Sec.5 the Chern-Simons  type
 geometry, the Lagrangian (and the problems in defining the measure of
integration), its gauge symmetries  and the  equation of motion of the
manifestly super-Poincare  invariant ten-dimensional gauge theory are
described. In Sec.6 the classical equation of motion is solved. The
solution is given in terms of  $N=1$ Majorana -Weyl superfield,
satisfying the manifestly supersymmetric gauge covariant Dirac equation.

\section{Gauge covariant BRST operator}
Consider a class of gauge theories for which the classical Hamiltonian is
proportional to the constraints. The path integral for the gauge -fixed
action of such theories can be presented in
 the form given by Batalin, Fradkin, Vilkovisky \cite{BFV}.
\begin{equation}
Z_{\P } = \int dP dQ \exp \left( i \int d\tau ( P \dot Q - H_{\P}) \right) \ ,
\label{PI} \end{equation}
where
\begin{equation}
H_{\P} = - \{\P , \O \} \ .
\label{HAM} \end{equation}
In  (\ref{HAM})  $\P  (P,Q) $ is the gauge fermion defining the
gauge-fixing  condition and $\O $ is the nilpotent BRST operator,
\\\begin{equation}
\O = \O(P,Q ) ;\     \O^2 =0  \ .
\end{equation}
The set of coordinates entering the gauge-fixed action  is denoted by $Q$
and the set of canonically conjugate momenta by $P$. Some of these
coordinates are original classical fields,
 some of them are the so-called ghost fields.

The gauge covariant BRST operator can be introduced into this path
integral by adding to the free BRST operator  $\O(P,Q) $  an
anticommuting connection field $A(Q)$, which is Lie algebra valued. The
background functional takes the form \begin{equation} Z_{\P }^{A}  = \int
dP dQ \exp \left( i \int d\tau ( P \dot Q + \{ \P,
 \O(P,Q ) + A(Q) \} ) \right) \ .
\label{BPI} \end{equation}
When the connection field $A$ is unconstrained, the background
functional can be shown to depend on the gauge fixing function $\P$. The
gauge symmetry transformation of the connection field $A$ can be
absorbed into the canonical change of variables and of the gauge fixing
function. \begin{eqnarray} (\O + A)^{\L}& =& e^{\L} (\O + A)e^{- \L} \ ,
\nonumber\\ \P^{\L} & =& e^{\L} \P  e^{- \L}\ .  \end{eqnarray} The
condition of the independence of the background functional (\ref{BPI}) on
the gauge fixing function $\P$ can be derived by performing the
infinitesimal change of integration variables given by (this is the
generalization of BFV theorem \cite{BFV}) \begin{eqnarray} Q^{\prime}
&=&Q + [Q , \O + A  ] \chi \ , \nonumber\\ P^{\prime} &=&P + [P , \O + A
] \chi\ , \end{eqnarray} where
\begin{equation}
\chi = i \int d\tau ( \P^{\prime} - \P)\ .
\end{equation}
The change in the measure of integration produces the term
\begin{equation}
\exp \left(-i \int \{\O + A , \P^\prime - \P \} d\tau \right) \ .
\end{equation}

It follows that
\begin{equation}
\frac{\partial Z^A_{\P}}{\partial\P } = < [(\O + A )^2 , \P ] >\ ,
\end{equation}
where the brackets $< \   >$  denote the functional integration with the
weight given in eq. (\ref{BPI}). Since the gauge-fixing function is
arbitrary we may conclude that the nilpotency of the gauge covariant BRST
operator  \begin{equation} (\O + A )^2  = 0 \label{nil}\end{equation}
 is the necessary and sufficient condition for the gauge independence of
the background functional (\ref{BPI}). Equation (\ref{nil}) is also known
to be the fundamental equation of the string field theory \cite{SFT},
following from the string field theory Lagrangian. \begin{equation}
\lagr_{SFT}=Tr A^{\dagger} (\half \O \star  A  +   \frac{1}{3} \star A \star
A)\ . \label{SFT}\end{equation} In the second quantized theory the
canonical momenta are realized as derivatives over the coordinate,
$P=\frac{\partial }{\partial Q }$. We denote by $d$ the differential
operator, related to the free BRST operator as \begin{equation}
 d (\frac{\partial}
{\partial   Q}, Q ) =\O (P=\frac{\partial }{\partial Q } , Q )\  ,
\end{equation}
and the covariant differential operator, related to gauge covariant BRST
operator introduced above, as \begin{equation}
D = d + A \  .
\end{equation}
The consistency condition of the first quantized theory, or the classical
equation of the second quantized theory take the geometrical form of the
Chern-Simons type equation for the curvature of the target space:
\begin{equation}\label{F}
F = D^2 = (d+A)^2 = 0\ .
\end{equation}

\section{D=10 superspace and chiral superfields}

According to the superparticle theory \cite{K}, the  flat D=10 superspace
can be characterized
 by the classical coordinates $X^\mu ,\theta_p^\a $, where $X^\mu$  is a
D=10 vector and  $\theta_p^\a $ are D=10 anticommuting Majorana-Weyl
spinors of positive chirality, $\a = 1, \dots, 16; \hskip 0.1 cm  p = 0,1,
\dots $. Our notations and conventions are those of \cite{SU}. The
important ones are the following. The spinors of positive chirality have
spinorial indices $\a$ up, the ones of negative chirality have them down.
The $\g$-matrices have their two indices both up or both down, and they
are symmetric.

 The  supersymmetry charges and covariant derivatives are defined as
follows. \begin{eqnarray}
q^p_\a &= &  {\frac{\partial}
{\partial \t_p^\a}}  + (\dsl \t_p)_\a \  ,\nonumber\\
d^p_\a &= & {\frac{\partial}
{\partial \t_p^\a}} -  (\dsl \t_p)_\a \  .
\end{eqnarray}
These charges and covariant derivatives  form the twisted $N=2(p+1)$
supersymmetry algebra,  \begin{eqnarray}
\{q^p_\a,q^l_\b\} &=&  + 2 \dsl_{\a\b} \delta  ^{pl} \ ,\nonumber\\
\{d^p_\a, d^l_\b \} &=& - 2 \dsl _{\a\b}\delta^{pl} \ , \nonumber\\
\{ d^p, q^l \} &=& 0 \ .
\label{tw}\end{eqnarray}
{}From the charges of the twisted $N=2(p+1)$ supersymmetry algebra
given in eqs. (\ref{tw}) we build two combinations,
\begin{eqnarray}
F^p_\a &= & \half( q^{p+1}_\a + d^p_\a) \  ,\nonumber\\
\tilde {F}^p_\a &= & \half (q^{p+1}_\a - d^p_\a)
 \  .
\end{eqnarray}
They form the following algebra:
\begin{eqnarray}
\{F^p_\a , F^l_\b\} &=& 0 \ , \nonumber\\
\{ \tilde {F}^p_\a , \tilde {F}^l_\b \} &=& 0 \ , \nonumber\\
\{ F^p_\a , \tilde {F}^l_\b  \} &=&  \dsl _{\a\b} \delta^{pl} \ .
\label{10FLAT}\end{eqnarray}
This algebra has Abelian subalgebras which remind the Abelian
subalgebras in D=4 where chiral and antichiral subalgebras exist,
\begin{eqnarray}
\{q^A , q^B\} &=& 0 \ , \nonumber\\
\{ q^{\dot{A}}, q^{\dot{B}} \} &=& 0 \ ,\nonumber\\
\{ q^A , q^{\dot{B}} \} &=& 2 \dsl ^{A\dot{B}} \ ,
\label{4FLAT}\end{eqnarray}
 $A,\dot A $ being 2-component spinor indices.
The treatment of D=10 flat superspace (\ref{10FLAT})  and gauge
theories in this superspace will have some analogies and some differences
with 4D flat superspace (\ref{4FLAT}) and gauge theories in it.

The irreducible representation of the algebra (\ref{10FLAT}) is
realized by putting the constraint which is analogous to the chiral
constraint in D=4. We can put the constraint (we will call it chiral
constraint by analogy) \begin{equation}
\tilde {F}^p_\a \hskip 0.1 cm  \Phi ( X^\mu , \t_l ) =( q^{p+1} - d^p )_\a
\hskip 0.1 cm   \Phi (X^\mu , \t_l ) =0\ .
\label{N} \end{equation}

The corresponding superfield can be real (as different from the chiral
4D superfield), since the constraint is real. The solution can be written
in the formal way as \begin{equation} \Phi ( X^\mu , \t_l )  = \prod \tilde
{F}^p_\a \hskip 0.1cm  \Psi ( X^\mu , \t_l ) \ , \label{SOL} \end{equation}
where $\Psi ( X^\mu , \t_q ) $ is an unconstrained superfield and the
product includes all $\tilde {F}^p_\a $. We will call the superfield
satisfying the constraint (\ref{N}) the chiral superfield. From all the
charges of the twisted  $N=2(p+1)$ supersymmetry algebra given in eq.
(\ref{tw}) only  \begin{equation} q^0 = \frac{\partial}{\partial \t_0} +
\dsl \t_0  \end{equation} commutes with the constraint $\tilde {F}^p_\a
$. Therefore the chiral superfield has the trivial dependence on
$\t_{p+1}$, even though it depends on all $\t_p,\  p=0,1,...$ and not only
on $\t_0$  as the the normal $N=1$ superfield  in $z = (X, \t_0)$
superspace does. The solution to the chirality constraint can be presented
by expanding the superfield in powers of $\t_1, \t_2, ...$, where the
coefficients of the expansion are normal $N=1$ superfields,
\begin{equation} \Phi ( X^\mu , \t_p ) = \Phi ( X^\mu , \t_0 )  + \Phi
^1_\a( X^\mu , \t_0 )  \t_1^\a +  \Phi ^{11}_{\a\b} ( X^\mu , \t_0 ) \t_1^\a
\t_1^\b + ... \  . \label{CH} \end{equation}

The chiral constraint allows to express all components of this superfield
through the first one and its covariant derivatives.  For example, the first
in the set of constraints (\ref{N}) is \begin{equation}
 \left(\frac{\partial}{\partial \t_0^\a} -
   (\dsl  \t_0)_\a \right)\Phi = \left( \frac{\partial}{\partial \t_{1} ^\a}
+ (\dsl \t_{1})_\a\right)
 \Phi ( X^\mu , \t_p ) \ .
\label{CS1} \end{equation}
When applied at $\t_1 = 0$, it gives
\begin{equation}
 \Phi ^1_\a ( X^\mu , \t_0 ) = d^0_\a
\Phi (X^\mu , \t_0 )
 \  .
 \end{equation}
Note, that $d^0_\a $ are the covariant derivatives for $N=1$ superfield
$\Phi
( X^\mu , \t_0 )$.

Alternatively, one may choose the constraint
\begin{equation}
 {F}^p_\a \hskip 0.1 cm  \tilde {\Phi} ( X^\mu , \t_l ) = ( q^{p+1} +d^p )
_\a \hskip 0.1 cm \tilde{ \Phi}( X^\mu , \t_l) =0\ .
\label{N1} \end{equation}
We will call this constraint and the corresponding superfield antichiral,
by analogy with D=4 superspace.

To describe the gauge theory, one proceeds in a standard way of
introducing connections for each  direction in flat superspace,
\begin{eqnarray}
{\cal F}^{p}_\a  &=&\half( Q^{p+1}_\a + D^p_\a ) =
\half (
q^{p+1}_\a +  A^{p+1}_\a  + d^p_\a + A^p_\a) \  ,\nonumber\\
\tilde {\cal F}^{p}_\a  &=& \half( Q^{p+1}_\a - D^p_\a ) =
\half (
q^{p+1}_\a +  A^{p+1}_\a  - d^p_\a - A^p_\a) \  .
\label{DER}\end{eqnarray}
One can rewrite this equations as follows:
\begin{eqnarray}
  D^p_\a &=& \frac{\partial}{\partial \t_p^\a} - \dsl  \t_p +
A^p_\a \  ,\nonumber\\
   Q^{p+1}_\a  &=&  \frac{\partial}{\partial \t_{p+1}^\a} + \dsl  \t_{p+1}
+  A^{p+1}_\a \  .
\label{COV}\end{eqnarray}

The gauge symmetry is introduced into the space by choosing  the
parameter of the gauge transformation to be a covariantly chiral
superfield, \begin{equation} \tilde {\cal F}^p_\a  \hskip 0.1 cm
\L(X^\mu , \t_l) =0\ .  \label{PAR} \end{equation} The integrability of
this constraint requires the vanishing of the   curvature,
\begin{equation}
\{\tilde {\cal F}^p_\a  , \tilde {\cal F}^q_\b \} = \tilde F^{p,q}_{\a \b}
=0\ . \label{RP} \end{equation}
One can choose the basis in which covariant chiral derivatives are  free
and  \begin{equation}
\tilde {\cal F}^p_\a  = \tilde { F}^p_\a\ ,\qquad
A^p_\a  -  A^{p+1}_\a =0\ ,\qquad
 {\cal F}^p_\a  =   F^p_\a +  A^p_\a(z)\ ,
\end{equation}
where $A^p_\a(z)$ is an unconstrained superfield. It is possible to
impose the set of constraints on torsion and curvature in the classical
superspace $z = (X^\m , \t_p)$  and to solve them to  describe the on shell
geometry.  In addition to eq. (\ref{RP}), we impose the conventional
constraint by expressing the vector connection through the spinor one:
\begin{equation} \half( \{\tilde {\cal F}^p_\a  , {\cal F}^q_\b \} +
\{\tilde {\cal F}^q_\b  , {\cal F}^p_\a \})
 =  \g^\mu _{\a\beta} \n_{\mu}\d^{p,q}\ .
  \end{equation}
The on shell constraint which is added to the geometry is
\begin{equation}
\{ {\cal F}^p_\a  , {\cal F}^q_\b \} = 0\ .
\label{RPON} \end{equation}
The solution of this last constraint requires the connection on shell to
be of the form \begin{equation}
A^p_\a(z) = e^V( F^p_\a e^{-V})\ ,
\end{equation}
where $V(z)$ is an unconstrained superfield. It is necessary to solve
Bianchi identities in the presence of all above mentioned constraints. We
have found that the solution requires the particular properties of
spinor-vector curvature, \begin{equation} [{\cal F}^p_\a , \n^{\a\g} ] =
\P ^\g \  . \label{spi}\end{equation}
This spinorial superfield must be covariantly antichiral and satisfy
gauge covariant Dirac equation. Thus the geometry of the superspace
described above is an alternative description of the on shell $D=10$
supersymmetric Yang-Mills theory. The difference with the usual $(X,
\t_0)$ superspace is the fact that it is related to the first quantized
superparticle action and that the geometry will be shown to be a solution
to the nilpotency condition of the gauge covariant BRST operator.

\section{First quantized superparticle}
The classical  action of the superparticle \cite{K} is
\begin{equation}
S^{cl} = P\dot X  + \sum_{p=0}^\infty \{\lambda^p \dot {\theta_p} +
T_a \psi^a\} \ ,
\label{CLD}\end{equation}
where $ \psi^a =  \{ g, \hskip 0.2truecm \zeta ^p,\hskip 0.2truecm\eta _p
\}$  are  the  Lagrange multipliers to the constraints\footnote {The second
ilk action in ref. \cite{SB} can be obtained from the action (\ref{CLD}) at
$\zeta ^p = 0, p = 1,2, ...$. This leads also to a different set of ghosts and
a
different BRST operator.}. The first-class constraints are
\begin{equation} T_a = \{ \half  P^2,\hskip 0.2truecm  K_p,\hskip
0.2truecm  F^p  \} \  . \label{T}\end{equation} Here we are using the
following notations: \begin{eqnarray} K_p &= & ( d^p + q^p )\Psl\
,\nonumber\\ F^p &= &\half ( d^p + q^{p+1})
 \  .
\end{eqnarray}
These constraints generate  local symmetries of the action:
re\-para\-metri\-za\-tion, $\k$- and $\xi$-symmetries, respectively.
Note, that the commutator of the fermionic $\k$- and $\xi$-symmetries
has  reparame\-tri\-zation in the right-hand side, \begin{equation} [ K_q
,\hskip 0.3truecm F^p ] = - \half P^2 (\delta _q^{p+1} - \delta ^p_q) \  .
\end{equation}
 In this sense the fermionic symmetries of the world-line are quite
fundamental and can be considered as the ``square root" of the Virasoro
constraint. The phase space of the  first quantized superparticle theory
\cite{BKV}--\cite{SU} in addition to classical coordinates $(X^\m ,
\t^\a_p)$ and their canonical momenta contains a set of ghost fields and
their canonical momenta. They are related to the local symmetries of the
action.

The gauge-fixed action in arbitrary gauge, including the light-cone
one and the free covariant one, has the following form \cite{K}:
\begin{equation}
\lagr_{gf}=P\dot X  + b\dot c
+\sum_{q=0}^{\infty }\sum_{p=q}^{\infty }\left[ \bar\lambda
^{p,q}\dot\theta _{p,q}+ \lambda _{p+1,q+1}
\dot{\bar\theta}^{p+1,q+1}\right]  +
\{\O_{BRST} , \Psi \} \ ,
\label{ASGF}\end{equation}
where  $\O_{BRST}$ is a nilpotent ghost number $+1$ BRST operator,
which will be presented below, and $\Psi$ is an arbitrary gauge fixing
function of ghost number $-1$.

In the gauge fixed-action (\ref{ASGF}) $b, c$ is a couple of
reparame\-tri\-zation ghosts. The spinors  (classical and ghost together)
form a spinorial multiplet of  $OSp(9.1/4)$ supergroup, consisting of
spinors of alternating Grassmann parity and chirality \cite{GH} -
\cite{BK}.

 We are using the notations of \cite{SU}, where the spinors of  $OSp
(9.1/4)$  are presented  using labels of SU(2) representations, e.g.
$\theta (j,m)$,  and/or by using their original labels as they come from
superstring  quantization,  where $\t_{p,q} = \t (j=p+q/2,\ m=p-q/2)$
and  ${\bar \t }^{p+1, q+1} = \t(j=p+q+1/2,\ m=q-p-1/2)$, $p\geq q \geq
0$.  The SU(2) label $m$ is related both to ghost number and to conformal
charge in string theory.

The nilpotent ghost number $+1$ BRST operator, which was found in
\cite{BKV}, is given, according to  \cite{K} -- \cite{SU}, by the following
expression:

\begin{equation}
\Omega =T_a(\phi ) c^a +\half b_a f^a{}_{bc}c^cc^b(-)^b
-b_{a_p} Z^{a_p}{}_{a_{p+1}}(\phi ) c^{a_{p+1}} +b_a b_{a_p}f^{a_pa}{}
_{a_{p+2}}c^{a_{p+2}} \ . \label{O}
\end{equation}

We assume a sum over $p=0,1,\ldots $ and $a_0 = a$. In (\ref{O})  $\phi $
are the classical fields of the action (\ref{CLD}). In constructing the BRST
operator we were using the formulation  given by Batalin, Fradkin and
Vilkovisky \cite{BFV}, in which $T$ are the original classical  first-class
constraints given in eq. (\ref{T}), $f$ are the structure constants of the
algebra and $Z$ are the zero modes. The BRST operator can be presented
in the following explicit for  \cite{BKV} - \cite{SU}.
 \begin{eqnarray} \Omega  &=&\half
cP^2 +\sum_{q=0}^{\infty }\sum_{p=q}^{\infty}
\bar \lambda ^{p,q}\left( \Psl \theta _{p+1,q}-2 \theta _{p+2,q}b \right)
\nonumber\\
&&+\sum_{q=1}^{\infty }\sum_{p=q+1}^{\infty}
\lambda _{p+1,q}\left( \Psl \bar \theta ^{p,q}-2 \bar
\theta^{p-1,q}b +\left(\bar \lambda ^{p+1,q}+\bar \lambda
^{p,q-1}\right)(-)^{\half(p-q+1)(p-q)} \right) \nonumber\\
&&+\sum_{p=1}^{\infty }\lambda _{p+1,p}\left( \Psl\bar
\theta ^{p,p}+\left(\bar
\lambda ^{p+1,p}+ \bar \lambda ^{p,p-1}\right)-2b\left(\theta
_{p,p}-\theta _{p-1,p-1}\right)\right)\nonumber\\
&&
+ \sum_{p=0}^{\infty }  \lambda _{p+1,p+1}\left( \bar \lambda
^{p+1,p+1}-\Psl \theta _{p+1,p+1}+2\theta _{p+2,p+1}b+\bar \lambda
^{p,p} + \Psl \theta _{p,p}-2\theta _{p+1,p}b\right) \  .
\label{newOmega} \end{eqnarray}

The BRST operator  commutes only with $N=1$ supersymmetry and  does
not commute with all of those extra supersymmetries which are not the
symmetries of the physical spectrum.

It was shown in \cite{BKV} that the cohomology of the nilpotent operator
(\ref{newOmega}) is given by the 8+8 supersymmetric Yang-Mills
multiplet.

\vskip 0.6 cm
In what follows we will use a version of the BRST operator related to the
one in eq. (\ref{newOmega}) by canonical transformation\footnote {I am
grateful to W.  Siegel for suggestion to represent the BRST operator, which
we have used for the second quantization, in this particular form.}
\begin{equation} \O^\prime = e^\Phi \O e^{-\Phi}\ ,
\label{PR}\end{equation} where  \begin{equation}
\Phi = b c_{p+1} \tilde F^p_\a \ ,
\end{equation}and
\begin{equation}
\tilde F^p_\a = \half ( d^p - q^{p+1}) \ .
\end{equation}
Consider now the functional given in eq. (\ref{BPI}). The set of classical
and quantum coordinates of the superparticle is defined above, the free
BRST operator is given in eq. (\ref{PR}). The gauge covariant BRST
operator will be introduced in a way that the connection field $A(Q)$ can
be absorbed in the change of variables in the path integral if the
connection is a pure gauge only.  Together with our choice of  the BRST
operator (\ref{PR}) we suggest a  particular choice of  coordinates  $Q$:
 \begin{equation}
Q = \{ Z , Y \}\  ,
\end{equation}
where
 \begin{eqnarray}
Z &=& \{ X^\m , \t^\a_p , b , c^\a_{p+1} \equiv \l_{p+1,p+1}\}\ ,
\nonumber\\
   Y &=& \{\bar \l^{p+1,q} , \bar \t ^{p+2,q+1}  \}\ ,\qquad p
\geq q \geq 0 \ .
\label{COO}\end{eqnarray}
With this choice of coordinates the free BRST operator $d$  is
\begin{equation}
 d =  \half P^2  \frac{\partial}{\partial b} +
 c^\a_{p+1} F^p_\a + b c^\a_{p+1} c^\b_{q+1} \{ \tilde F^p_\a ,
  F^q_\b \} + d_Y \frac{\partial}{\partial Y}\ .
 \label{fulld}\end{equation}
In the general case the consistency condition for the background field
$A(Q)$ is given by the nilpotency condition of the covariant BRST
operator and the gauge symmetry of the target space is defined by the
transformation \begin{equation}
d + A^\prime =  e^{\L} (d + A) e^{- \L}\ ,
\end{equation}
where $d$ is given above and the parameter is some general function of
coordinates $Q$. One has to be much more specific if one wants to build
not only the geometry of the background of the first quantized theory but
also to understand the way to perform the second quantization of the
superparticle theory.

 \section{Non-Abelian D=10 Chern-Simons theory}

 The quadratic second quantized action for the superparticle was
analysed in \cite{PAS}. This action describes the Abelian
supersymmetric gauge theory in D=10, where it is a free theory, since the
Majorana spinors do not interact with the Abelian vector field.
\begin{equation}
 S = \half  \int dQ A(Q) d(\frac{\partial}{\partial Q}, Q) A(Q) \ .
\label{ABS}\end{equation}
Here $ d$ is the BRST operator given in eq. (\ref{fulld}) and $A$ is the
anticommuting connection.  With our choice of  coordinates and momenta
all terms in $d$ are at least linear in derivatives for the BRST operator
\cite{K} -- \cite{SU} of
 the superparticle action \cite{K}.  This means that $d$ is a differential
operator  and the rules of partial integration can be applied.

This action avoids the old problem of $D=10$ supersymmetry. There
exists the no-go theorem which says that one cannot construct a
supersymmetric Lagrangian for  this theory  if it contains a  {\it finite
number of auxiliary fields}.  In our case the  off-shell connection $A$ is
a general superfield in ($X, \t_p$) space containing an {\it infinite}
number of  components.

There is a problem, however, in defining the measure of integration over
the coordinates of the superparticle in (\ref{ABS}), since they include
commuting spinors of a given ghost number. We hope that this problem
will be solved and the action (\ref{ABS}) will be used for the construction
of the off-shell theory. This would be necessary for consistency of the
second quantized theory (\ref{ABS}) and the first quantized
superparticle theory. Indeed, if the measure of integration in (\ref{ABS})
is properly defined, then the variation of this action with respect to the
field $A$  leads to the same CS equation as the one derived from the
condition of consistency of the first quantized superparticle theory in
the background field $A$.

The equation of motion following from
(\ref{ABS}) is \begin{equation} d(\frac{\partial}{\partial Q}, Q) A(Q) =
0 \ . \label{ABEQ}\end{equation}
The action (\ref{ABS}) and the equation of motion (\ref{ABEQ}), have a
gauge  symmetry under the following transformation of a connection field
$A(Q)$, \begin{equation}
\d A(Q) = d(\frac{\partial}{\partial Q}, Q) \Lambda (Q)  \ ,
\label {SY}\end{equation}
since $d$ is a fermionic nilpotent operator, satisfying the equation
\begin{equation}
d^2 = \half \{d , d\} = 0\ .
\end{equation}
It was shown in \cite{BKV} that it is sufficient to consider the case  $P^2
= 0$, when looking for the non-trivial solution of the cohomology. In
addition to that we were looking for the solutions of the eq.  (\ref{ABEQ})
which has the property of being independent on the coordinates, denoted
by $Y$, i.e. on  $(\bar \l^{p+1,q} , \bar \t ^{p+2,q+1} )$. It has been found
under these  conditions that the solution of equation  (\ref{ABEQ})
reproduces $D=10$ supersymmetric QED.

The complications which arise in the non-Abelian case are related to the
fact that  the full BRST operator in equation  (\ref{fulld}) has some terms
which are quadratic or even cubic in derivatives\footnote{This
observation has been made by M. Peskin.}. If we are going to use the string
field type action \begin{equation}
 S = \half  \int dZ ( A d_1 A +
\frac{2}{3} A^3 ) \ ,
\label{NABS}\end{equation}
we are allowed to use only the part of $d$ which is linear in derivatives
to be able to prove the non-Abelian gauge symmetry of the action
(\ref{NABS}). Also we need to consider a smaller set of coordinates, which
includes only $Z$-coordinates, see eq. (\ref{COO}). Fortunately, as it
follows from the analysis of the Abelian theory, the corresponding part of
$d$ \begin{equation}
 d_1 =
 c^\a_{p+1} F^p_\a + b c^\a_{p+1} c^\b_{q+1} \{ \tilde F^p_\a ,
  F^q_\b \}
 \label{smalld}\end{equation}
is indeed responsible for the non-trivial solution to cohomology. The
equation of motion following from  the action (\ref{NABS}), (\ref{smalld})
has the CS form \begin{equation} F_1 = D_1^2 = (d_1 + A(Z))^2 = 0\ .
\label{chern}\end{equation}
Thus, from the second quantized theory (\ref{NABS}) we have derived  (at
least formally until the integration over the coordinates of the
superparticle will be better understood) the same equation which in a
more general form with a full $d$ and with an arbitrary choice of all
coordinates versus momenta has been derived from the consistency of the
first quantized superparticle theory in a background field $A$.

\section{Solution to CS equation of motion}
The cohomology analysis of the free BRST operator performed in
\cite{BKV} has shown that the only   non-trivial solution describes the
$D=10$ supersymmetric multiplet.  Now we are going to demonstrate that
the solution of the simple CS equation $D^2 = 0$ (\ref{F}), which is the
condition of the nilpotency of the gauge covariant BRST operator $D$,
gives a full non-linear description of the ten-dimensional
supersymmetric Yang-Mills theory.

 The solution we are looking for will not depend on $Y$. Therefore, to
find a non-trivial  solution it will be sufficient to solve a simpler
equation (\ref{chern}).

To solve this equation, we consider the target superspace
 \begin{equation}
Z  = \{ X^\m , \t^\a_p , b , c^\a_{p+1} \}\ ,\qquad p= 0,1, \dots ,\infty \ .
\label{QS}\end{equation}
In addition to classical coordinates $z =  ( X^\m , \t^\a_p)$, it has also
the anticommuting reparametri\-zation antighost $b$, with the ghost
charge $-1$, and  commuting $\xi$-symmetry ghosts $\l^\a_{p+1,p+1}
\equiv c^\a_{p+1}$, each with the ghost charge $+1$.

The anticommuting connection is $A = A_0 + b A_1$, where $A_0$ has
ghost number +1 and therefore is linear in $c^\a_{p+1}$ and $A_1$ has
ghost number +2 and is quadratic in $c^\a_{p+1}$.
 \begin{equation}
A(Z)  = c^{\a}_{p+1} A_\a^p (z) +  b  c^{\a}_{p+1}  c^{\b}_{q+1}
A^{pq}_{\a\b} (z)\ .
\label{C}\end{equation}
 $A_\a^p (z)$ and $A^{pq}_{\a\b} (z)$ are two arbitrary independent
functions on classical superspace $z$. The covariant derivative is
\begin{equation}
D_1 = c^{\a}_{p+1} {\cal F}^p_\a + b c^\a_{p+1} c^\b_{q+1}
(\{ \tilde F^p_\a, F^q_\b \} + A^{p,q}_{\a\b})\ .
\end{equation}
Under the gauge transformations it transforms as follows:
\begin{equation}
D^\prime_1 = e^\L D_1 e^{-\L}
\ ,
\label{gauge}\end{equation}
where the parameter of the gauge transformation $\L$ has zero ghost
charge and has the form   \begin{equation}
\L(Z) = \L_0(z) + b c^\a _{p+1} \L_\a^{p}(z)
\ .
\end{equation}
We can partially use the gauge freedom of $D_1$ by expressing $A_1$
through $A_0$ in the solutions,  \begin{equation}
A_1 =  c^\a_{p+1} \tilde F^p_\a A_0\ .
\label{CONS}\end{equation}
In this gauge our covariant derivative takes the form
\begin{equation}
D_1 =  c^{\a}_{p+1} {\cal F}^p_\a + b
 c^\a_{p+1} c^\b_{q+1} \{ \tilde F^p_\a, {\cal F}^q_\b \}
\end{equation}
The remaining gauge symmetry is given by eq. (\ref{gauge})   with the
only remaining parameter of transformation $\L_0$ being a chiral
superfield, satisfying the equation \begin{equation}
 \tilde F^p_\a \L_0 = 0\ .
\label{CHIR}\end{equation}

The CS equation in these gauge is
\begin{equation}
F_1 = D_1^2 =  c^\a_{p+1} c^\b_{q+1} \{ {\cal F}^p_\a, {\cal F}^q_\b \} +
    b  c^\a_{p+1} c^\b_{q+1}c^\g_{l+1}[ {\cal F}^p_\a,    \{ \tilde F^q_\b,
{\cal F}^l_\g \}] = 0 \ . \end{equation}
We have found that the solution of this equation coincides with the
solution of Bianchi identities in the classical superspace with the
constraints, as given in Sec.3. In particular, we have to require that in
this gauge  \begin{equation}
{\cal F}^p_\a =  e^{V(z)} F^p_\a e^{-V(z)}\ ,
\end{equation}
where $V$ is a general function of $z$. In addition to that the gauge
covariant  curvature in the spinor-vector direction build up from $V$
according to equation (\ref{spi})
must satisfy a list of
properties. In particular it is a covariantly antichiral superfield, for
which the dependence on $\t_1, \t_2, ...$ is defined by the dependence on
$\t_0$. This kind of a superfield is very close to the normal $N=1$
superfield, depending only on $\t_0$. This superfield satisfies the Dirac
equation and describes the on shell content of the non-Abelian
supersymmetric gauge theory in $D=10$: \begin{eqnarray}
\g^{\m} \n_{\m} \Psi (X, \t _p) & = & 0\ ,\nonumber\\
{\cal F}_\a^p \Psi ^\b (X, \t _q) & = & 0\ ,\nonumber\\
\tilde F_\a^p \Psi ^\b (X, \t _q) & = & \g_\a^{\m \n \b} F_{\m\nu}\ .
\end{eqnarray}

Thus, the nilpotent gauge covariant  derivative $D_1$ in this gauge is

\begin{equation}
D_1 =  e^{V(z)}  c^\a_{p+1}F^p_\a e^{-V(z)} +  bc^\a_{p+1} c^\b_{q+1}\n
_{\a\b }\d^{p,q} \ , \end{equation}
where the general superfield $V$ defines the covariantly antichiral
superfield $\Psi$ according to eq. (\ref{spi}) and $\Psi$ has the
properties presented above. The dependence on $V$ and specifically on
gauge covariant combination $\Psi$ cannot be gauged away in this gauge,
since the only remaining symmetry is the one with the chiral parameter.
In this sense it is a non-trivial solution to the cohomology equation which
describes ten-dimensional supersymmetry.

We hope that this construction can be generalized to incorporate    D=10
supergravity interacting with the supersymmetric Yang-Mills system.
This would be an important step towards  understanding  the  superstring
geometry.

\vskip 0.6 cm

I would like to thank  M. Peskin for the useful discussions and
participation at the early stage of the investigation. I wish to express my
gratitude to  E. Bergshoeff,  W. Siegel and A. Van Proeyen
 for the very productive discussions of the issues related to second
quantization of the superparticle.


\vfill
\newpage




\begin{thebibliography}{50}
\bibitem{NIL} B.E.W. Nilsson,  preprint G\" oteborg-81-6 (1981).
 \bibitem{BKV} E.A. Bergshoeff, R. E. Kallosh and A. Van Proeyen,  Phys.
Lett. {\bf B251} (1990) 128.
\bibitem{K} R. E. Kallosh, Phys. Lett. {\bf B251} (1990) 134.
\bibitem{SU} E. Bergshoeff, R. E. Kallosh and A. Van Proeyen,  {\it
Superparticle Actions  and Gauge Fixings}, CERN preprint TH.6020/91
and Stanford University  preprint SU-ITP-888 (1991), submitted to Class.
Quant. Grav. \bibitem{SFT} E. Witten, Nucl. Phys. {\bf B268} (1986)
253.\\ For a review of the string field theory see: W. Siegel, {\bf
Introduction to String  Field Theory} (World Scientific, Singapore, 1988);
C. Thorn, Phys.Rep.  {\bf  175} (1989) 1.
\bibitem{BM} R. E. Kallosh, Phys. Lett.{\bf B195} (1987) 369;
W. Siegel, in: {\bf Strings 89} (World Scientific,
Singapore, 1990)  p. 211; U. Lindstr\" om, P. van Nieuwenhuizen,     M.
Ro\v cek,  W. Siegel and A. van de Ven, Phys. Lett. {\bf B224} (1989) 285;
 M.B. Green and C.M. Hull, in: {\bf Strings 89} (World Scientific,
Singapore, 1990)  p. 478;
R.E. Kallosh, Phys. Lett. {\bf B224} (1989) 273;
Phys. Lett. {\bf B225} (1989) 49;
M. Green and C. Hull, Phys. Lett. {\bf B225} (1989) 57;
J. Gates, M. Grisaru, U. Lindstr\" om, P. van Nieuwenhuizen, M. Ro\v cek,
W.Siegel and A. van de Ven, Phys. Lett. {\bf B225} (1989) 44;
E.A. Bergshoeff and R. E. Kallosh,  Nucl. Phys. {\bf B333} (1990) 605 .
M.B. Green and C.M. Hull, Nucl. Phys. {\bf B344} (1990) 115;
M.B. Green and C.M. Hull, Mod. Phys. Lett. {\bf A5} (1990) 1399
and contribution to the conference "Strings '90".
\bibitem{SB} A. Mikovi\'c, M. Ro\v{c}ek, W. Siegel, P. van
Nieuwenhuizen, J. Yamron and A.E. van de Ven, Phys. Lett. {\bf B235}
(1990) 106;
F. E{\ss}ler, E. Laenen, W. Siegel and J.P. Yamron, Phys. Lett. {\bf  B254 }
(1991)  411; F. E{\ss}ler, M. Hatsuda, E. Laenen, W. Siegel, J.P. Yamron, T.
Kimura and A. Mikovi\'c, preprint ITP-SB-90-77.
\bibitem{BFV}  E.S. Fradkin and G. Vilkovisky, Phys. Lett. {\bf 55B}
(1975) 224; I.A. Batalin and E.S. Fradkin, Phys. Lett. {\bf B122} (1983)
157; I. Batalin and G. Vilkovisky,  Phys. Lett. {\bf 69B} (1977) 309; Phys.
Rev. {\bf D28} (1983) 2567;  M. Henneaux, Phys. Rep. {\bf 126} (1985) 1.
\bibitem{GH} M.B. Green and C.M. Hull, Phys. Lett. {\bf B229} (1989) 215.
 \bibitem{BAK} I. Bars and R. E. Kallosh, Phys. Lett. {\bf B233} (1989)
117.  \bibitem{BK} E.A. Bergshoeff and R.E. Kallosh, Phys. Lett. {\bf
B240} (1990) 105.  \bibitem{PAS} R. Kallosh, in: Proceedings of the
PASCOS conference, 1991.







\end{thebibliography}
\end{document}